\documentclass{kapproc} % Computer Modern font calls

\usepackage{procps} 

\usepackage{graphicx}

\upperandlowercase

 \let\footnote\savefootnote

\normallatexbib %will produce bibliography entries as shown in the
\newcommand{\ion}[2]{#1$~${\scshape{#2}}}

\newcommand{\ha}{H$\alpha$}
\newcommand{\hb}{H$\beta$}
\newcommand{\sii}{\ion{S}{II}}
\newcommand{\nii}{\ion{N}{II}}
\newcommand{\oi}{\ion{O}{I}}
\newcommand{\oiii}{\ion{O}{III}}
\newcommand{\hei}{\ion{He}{I}}
\newcommand{\hi}{\ion{H}{I}}
\newcommand{\hii}{\ion{H}{II}}
\newcommand{\niib}{[NII]$~\lambda$5755}

\newcommand{\dg}{$^\circ$}
\renewcommand\th{\thinspace}
\newcommand\kms{\ifmmode{\rm km\th s^{-1}}\else km\th s$^{-1}$\fi}

\begin{document}

\articletitle
{The WHAM Northern Sky Survey and the Nature of the Warm Ionized Medium in the Galaxy}

\chaptitlerunninghead{The WHAM Survey} % Shorter running head title.

\author{G. J. Madsen}%\altaffilmark{1}}

\affil{University of Wisconsin -- Madison }

\begin{abstract}

The Wisconsin H-Alpha Mapper (WHAM) has completed a velocity-resolved map of diffuse \ha\ emission of the entire northern sky, providing the first comprehensive picture of both the distribution and kinematics of diffuse ionized gas in the Galaxy.  WHAM continues to advance our understanding of the physical conditions of the warm ionized medium through observations of other optical emission lines throughout the Galactic disk and halo.  We discuss some highlights from the survey, including an optical window into the inner Galaxy and the relationship between \hi\ and \hii\ in the diffuse ISM.

\end{abstract}

\section{Introduction}

The warm ionized medium (WIM) is now recognized as a significant component of the
interstellar medium. Several studies over the past
few decades have revealed the presence of diffuse ($n_e \sim
0.1~$cm$^{-3}$), warm 
($T_e \sim 10^4 K$) ionized gas spread throughout the
Galaxy. Historically, this phase 
of the ISM  was thought to be confined primarily to classical \hii\
regions and planetary nebulae. However, the spectral characteristics
of the 
Galactic synchotron emission (Hoyle \& Ellis 1963), the discovery of
pulsars and their dispersion measures (Guelin 1974), and the
detection of faint optical emission lines (Reynolds et al 1974) 
all lead to the notion that a widespread, diffuse layer of
\hii\ permeates the Galaxy.
This gas is now known to occupy a significant fraction of the Galaxy (a volume filling
fraction $\sim$ 20\%) and
account for most (90\%) of the photoionized gas in the solar
neighborhood (d $\le 2-3$~kpc) (Reynolds 1991; Taylor \& Cordes 1993; Mitra, this conference).
With a scale height of $\sim$1 kpc (Haffner, Reynolds, \& Tufte
1999), and a 
column density perpendicular to the plane of $N_{H^+} \sim 1/3~N_{HI}$
(Reynolds 1989), this  
material plays a crucial role in our understanding of the physical
conditions and dynamics of the ISM in general. The presence of diffuse ionized gas in external galaxies is also now firmly established (Dettmar, this conference).

Despite the fact that it is a significant component of the
interstellar medium, the origin and
physical conditions within the WIM remain poorly understood. Questions
such as how the WIM is ionized, what is its source of heating, and how its
structures are formed have yet to be fully answered. The Lyman
continuum radiation from OB stars is the only known source with
sufficient power to ionize the  
WIM (see Beckman, this conference); however, it is not understood how this radiation, originating
from widely separated, discrete regions near the midplane, is able to
penetrate the ubiquitous neutral hydrogen to produce this
widely spread H$^+$ within the disk and halo. 
The energy from a variety of sources such as 
supernovae, hot white dwarf stars, turbulent mixing layers, and magnetic reconnection may also contribute, but they appear incapable of producing most of the ionization (Reynolds
1990; Slavin, this conference). 

Even though the primary source of ionization is believed to be
O stars, the temperature and ionization conditions within the diffuse
ionized gas appear to differ significantly from conditions within
classical O star \hii\ regions.  For example, anomalously strong
[\sii]$~\lambda6716$/\ha\ and [\nii]$~\lambda6583$/\ha, and weak
[\oiii]$~\lambda5007$/\ha\ emission line ratios (compared to the
bright, classical \hii\ regions) indicate a low state of excitation
with few ions present that require ionization energies greater than 23
eV (Haffner et al 1999; Rand 1997; Wood, this conference).  This is consistent with the low
ionization fraction of helium, at least for the helium near the
midplane, implying that the spectrum of the diffuse interstellar radiation
field that ionizes the hydrogen is significantly softer than that from
the average Galactic O star population (Reynolds \& Tufte 1995, Heiles et al 1996).

Photoionization models also fail to account fully for 
observations of line ratios among some of the other emission lines.
For example, the models do not
explain the very large increases in [\sii]/\ha\ and [\nii]/\ha\ 
(accompanied by an increase in [\oiii]) with distance from the
midplane or the observed constancy of [\sii]/[\nii] (see Reynolds et al 1999, Haffner et al 1999, Collins \& Rand 2001).
There is also growing evidence
that the WIM requires an additional heating source other than
photoionization, as 
revealed by observations of [\sii], [\nii], and \niib~(Reynolds et al 2001,
Reynolds, Haffner, \& Tufte 1999). Photoelectric heating by grains,
dissipation of turbulence, damping of MHD waves, and cosmic 
ray interactions have all been proposed as supplemental heating
sources. 

The warm ionized medium clearly has an important bearing on our understanding of the composition and structure of the interstellar medium and the processes of ionization and heating in the Galactic disk and halo.

 \section{The WHAM Northern Sky Survey}

The Wisconsin \ha\ Mapper has mapped the entire northern sky ($\delta \ge -30$\dg) in the
brightest optical emission line of the WIM, Balmer
\ha\ (Haffner et al 2003). 
With its dual-etalon Fabry-Perot design, one degree
diameter field of view, 12 \kms\ spectral resolution within a 200 \kms\ spectral
window, and unprecedented sensitivity of 0.1 Rayleigh, WHAM has produced
maps of the \hii\ comparable to the 21 cm surveys of the \hi.
These maps reveal for the first time the detailed structure of the
WIM, including long  
filaments, loops, ``worms'', and ``point sources'' superposed on a diffuse background (Fig.\ 1).  
Whereas other \ha\ surveys 
(eg. Gaustad et al 2001, Parker \& Phillipps 1998, Dennison et al 1998) 
are limited in sensitivity, spatial
coverage, or velocity resolution, the WHAM Northern Sky Survey (WHAM-NSS), although having lower angular resolution, has
provided a complete view of both the distribution and kinematics of the
WIM over the entire northern sky. 
WHAM is now poised to play an equally significant role
characterizing the physical conditions within the gas, exploring the
relationship of this warm ionized phase to the other principal phases
of the medium, and investigating sources of ionization and heating
within the Galactic disk and halo.
 
 \section{Exploring the Heterogeneous Nature of the WIM}

The spectacular structure revealed by the new detailed view of the WIM from the WHAM-NSS naturally leads to a number of interesting questions.
For example, how do HII regions differ from structures in the WIM and the diffuse WIM?
How can the variations in the line ratios be explained?
Why is the \hii\ so widespread in the disk and halo?
Is the diffuse WIM a superposition of filaments, loops, etc?
To help answer these questions, we have begun examining several interesting \ha\ features through observations in other optical emission lines. The maps in Figures 2 and 3 show regions of the WIM that sample classical \hii\ regions and shells, large-scale filaments with no associated ionizing sources, and the more truly diffuse WIM.
By analyzing the relative strengths, line ratios, and line widths of \ha, \hei, [\nii], [\sii], and [\oiii] of these features, the temperature and ionization conditions of the gas can be inferred (Haffner et al 1999, Reynolds et al 1999). 
For example, diagnostic plots of [\nii]/\ha\ vs. [\sii]/\ha, can be used to compare the derived electron temperature of the gas and ionization state of sulfur among the different WIM environments (Fig.\ 1). Populating plots like these with observations toward a large number of directions may help to classify these structures and understand their origin.  Observations of lines from other ions such as [\oiii] will further constrain the ionization conditions and lead to some insight as to how these features might be related to one another (Madsen et al 2003, in prep).

 \begin{figure}[t]
  \includegraphics[scale=0.64]{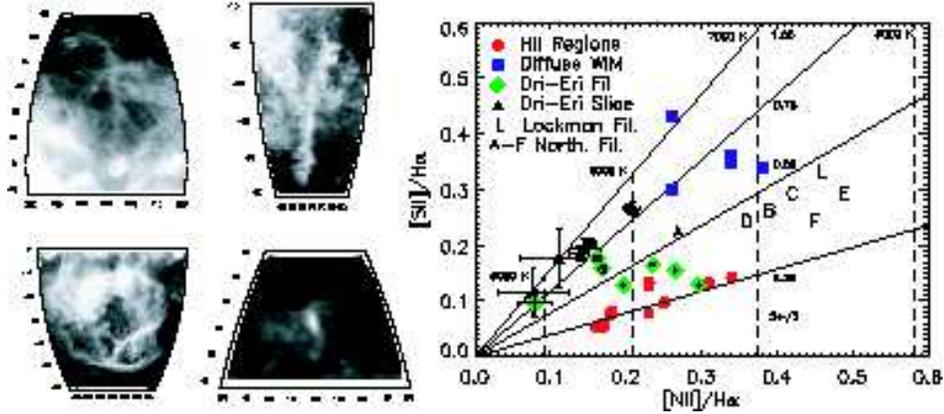}
  \caption{Samples of \ha\ emission structures from the WHAM-NSS with a diagnostic line ratio plot. The maps show the variety of the \ha\ emitting features in the WIM, with only some having known sources of ionization.   Observations of the relative line strengths of [\sii], [\nii], and \ha\ toward structures like these can be used to constrain their physical conditions. Using [\nii]/\ha\ as a measure of temperature, and [\nii]/[\sii] as a measure of S$^+$/S, the plot on the right can be used to classify these features and may lead to an understanding of their origin.}
\end{figure}

 \section{An Optical Window into the Inner Galaxy}

One of the more surprising results from the WHAM-NSS is the visibility of significant amounts of \ha\ emission at velocities exceeding +100~\kms~(LSR) toward the inner Galaxy ($20^\circ \le l \le 40^\circ, -5^\circ \le  b \le 0^\circ$), implying the detection of ionized gas at large distances from the Sun.  
The velocity coverage of these observations have been extended out to +150 \kms, beyond the tangent point velocities (see Fig.\ 2). These new observations have confirmed the Survey's detection of faint emission at high velocities from this very distant gas (Madsen et al 2003, in prep). In a 5\dg\ by 5\dg\ area, we detect diffuse optical emission a few degrees off the Galactic plane at velocities that place the gas near the tangent point at kinematic distances of 7-8 kpc, assuming the gas participates in Galactic rotation.  
\hb\ observations of this same region show a monotonic increase of the \ha/\hb\ ratio with increasing velocity (or distance). This suggests that 1) interstellar dust is affecting this emission, and 2) that the emission at higher velocities is traversing a longer path containing dust, confirming that the higher velocity emission originates at larger heliocentric distances and is probing the inner parts of the Galaxy. 
With a higher star-formation rate, larger gravitational pressure, and stronger UV flux, the temperature and ionization conditions of diffuse ionized gas in the inner Galaxy are likely to vary significantly from what is found nearby.  
However, studies of the WIM close to the midplane where the ionizing stars are located, were thought to be prohibitive due to optical extinction.
This optical window provides a unique opportunity to explore, for the first time, the properties of diffuse ionized gas in the central regions of the Galaxy using the same optical emission line diagnostic techniques to examine the WIM in the solar neighborhood.

\begin{figure}[t]
  \includegraphics[scale=0.64]{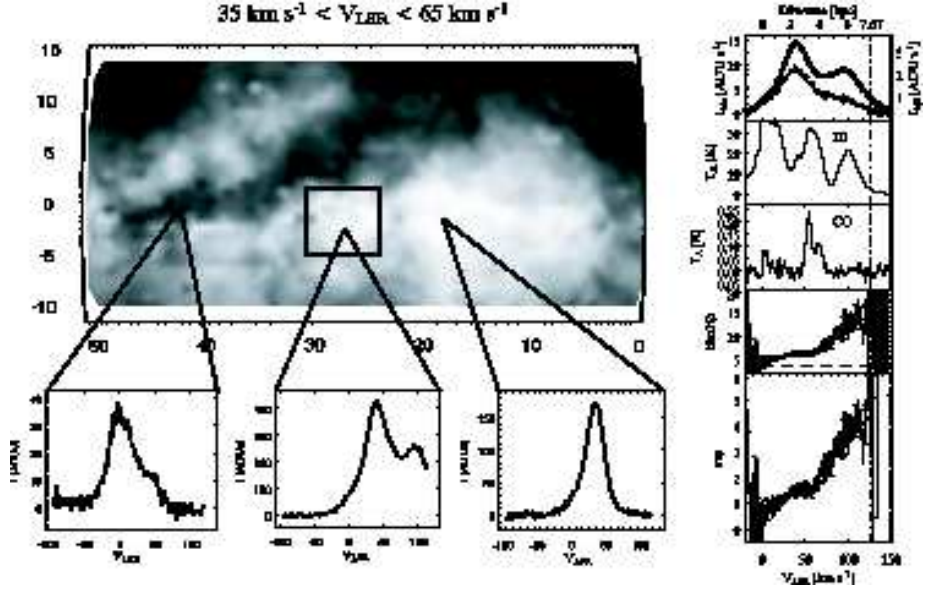}
  \caption{\ha\ map and spectra from the WHAM-NSS toward the inner Galaxy.  The large dark feature in the middle of the map is the Aquila rift, a nearby dust cloud ($d \approx 250$ pc) that is obscuring the \ha\ emission behind it with $v_{\rm{LSR}}$ > 25 \kms.
The three diagrams below the map show the \ha\ spectra toward the indicated directions, revealing the change in the obscuration of emission from the local neighborhood (0 \kms), the Sagittarius arm (+50~\kms), and the Scutum arm (+80 \kms). The black box is the low extinction window with emission out to the tangent point velocity. 
The plot on the right shows an \ha, \hb, \hi, and CO spectrum toward this window out to the tangent point velocity, denoted by a vertical dashed line. The inferred attenuation ($\tau_{\rm{H}\beta}$) as a function of velocity (distance), derived from \ha/\hb, is shown at the bottom.}
\end{figure}
 
  \section{The Relationship between \hi\ and \hii}

 \begin{figure}[t]
  \includegraphics[scale=0.64]{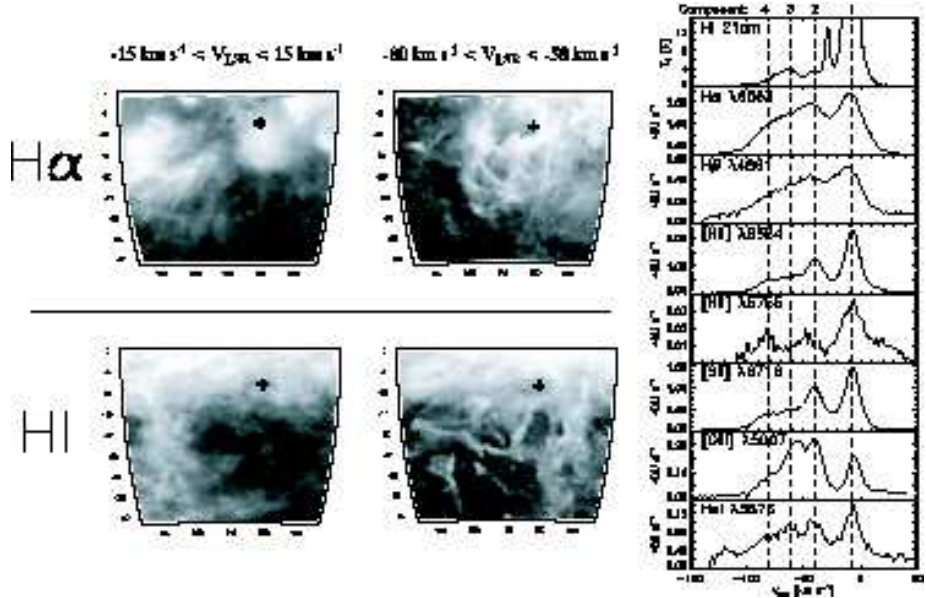}
  \caption{\ha\ and \hi\ maps with multi-wavelength spectra. The upper \ha\ maps are from the WHAM-NSS with the lower \hi\ maps from the LDS, and they span two different velocity intervals sampling the local ISM and the Perseus spiral arm. Note the lack of correspondence between \ha\ and \hi\ features from nearby gas on the maps on the left, whereas a correspondence between some filaments and clouds begin to emerge from the distant gas in the maps on the right.  The spectra on the far right are of different emission lines toward the direction indicated by the plus sign.  Four distinct velocity components in the ionized gas are identified, with significant variations in the line ratios from one component to another.}
\end{figure}

The relationship between neutral atomic hydrogen and warm ionized hydrogen, the two principal states of interstellar gas above the Galactic plane ($b~>$~5\dg), is not well known, and there are a number of possibilities. The \hii\ may be confined to the outer envelopes of \hi\ clouds embedded in a low-density, hot ionized medium (McKee \& Ostriker 1977).  The \hii\ may be the fully ionized component of a widespread warm neutral medium (Miller \& Cox 1993). The \hii\ may be well mixed with \hi\ in partially ionized clouds (Spitzer \& Fitzpatrick 1993). 
Several detailed studies of relatively high-density regions of the ISM near the Galactic midplane (e.g. individual \hii\ regions and supernova remnants) have established relationships between these phases on a small scale. However, the relationship, if any, between \hii\ and \hi\ in the large-scale diffuse ISM has yet to be established observationally.  
With an \ha\ spectral map of the whole northern sky now complete,  as well as comparable all-sky \hi\ surveys such as the Leiden-Dwingeloo Survey (LDS, Hartmann \& Burton 1997) and the \hi\ Parkes All Sky Survey (Barnes et al 2001), it is now possible to address this question to determine how \hii\ and \hi\ are related as a whole in the diffuse ISM. 

Preliminary studies of this relationship highlight the complexities of such an effort.  When integrated over all velocities, the total intensity maps of \ha\ from the WHAM-NSS, and \hi\ from the LDS, bear little or no morphological  resemblance to one another (see Fig.\ 3).
However, the total intensity maps are dominated by emission from cold \hi\ (narrow line-width) clouds at velocities near the local standard of rest. These cold clouds may be residing inside warm \hi\ envelopes, and hence have little associated ionized hydrogen. 
By moving out to intermediate velocities, isolating emission from warm gas above spiral arms, a morphological correspondence begins to emerge, suggesting that at least some of the \hii\ is somehow related to the \hi\ clouds away from the Galactic plane. 
Examining the \ha\ and \hi\ spectra and the morphology toward a large region of the Perseus arm (Fig.\ 3) shows that  there is 1) a good spatial correlation between individual \hii\ and \hi\ filaments and 2) a good correlation of \hii\ with {\it{warm}} \hi\ emission components. There is little or no correlation between the  \ha\ emission and the cold \hi\ emission components, as expected. In addition, there is little or no correlation between the {\it{strength}} of the \hii\ and warm \hi\ emission, as expected since the hydrogen in the WIM is likely photoionized.  

These observations do not establish if the \hii\ and warm \hi\ are well mixed, or separated by an \hi--\hii\ cloud boundary.  An ideal program would be to identify a population of well-defined, intermediate-velocity, isolated warm \hi\ clouds and look for limb-brightened associated \ha\ emission on the edge of the \hi\ clouds.  However, the relatively low densities ($n \approx 0.3$ cm$^{-3}$) and incident ionizing flux ($\Phi \approx 2.0\times10^6~$photons~cm$^{-2}$~s$^{-1}$) of these clouds dilute this limb brightening effect, resulting in relatively large ionized layers ($l\approx 50\ $pc) with low contrast (Madsen et al 2001).
Alternatively, constraints on the ionization fraction of hydrogen of the clouds can be made through observations of [\oi]~$\lambda$6300/\ha\  (Reynolds et al 2001). A future observational program searching for changes in the ionization fraction across the faces of \hi\ clouds through [\oi]\ spectra will help to assess the relationship between \hi\ and \hii.

 \begin{acknowledgments}
WHAM continues to explore diffuse ionized gas through the generous support of the National Science Foundation, through grant AST 02-04973. G.J.M. also acknowleges support from the Wisconsin Space Grant Consortium. 
\end{acknowledgments}

\section{Discussion}

\noindent
{\it J. Slavin:} Can you see a dust shadow of \ha\ toward a nearby dusty cloud? It would be useful for locating the WIM emission. \\

\noindent
{\it Madsen:} Yes, the Aquila rift is a nearby dust cloud (d $\approx$ 250 pc) toward the inner Galaxy that we see an \ha\ shadow against (see Fig.\ 2).  However, this method is limited by the relatively small number of well-defined, large ($> 1$\dg) dust clouds of known distance.  \\

\noindent
{\it R. Sankrit:} In your plot of [\sii]/\ha\ vs. [\nii]/\ha, did you include B-star \hii\ regions? If not, will the eventually be included on it? \\

\noindent
{\it Madsen:} Currently there are no B-star \hii\ regions on that plot, but only because we have not yet reduced those data to include them. Increasing the number of observations toward different environments on that plot will help in understanding how they may be related to each other. \\

\noindent
{\it J. Hester:} How does the structure along the newly discovered window compare with the structure toward Baade's window? \\

\noindent
{\it Madsen:} They are both very similar, however Baade's window is a much smaller  region of the Galaxy ($\approx$ 1\dg).  Both are likely to be chance superpositions of the voids between the clumpy dust structures in the Galactic plane.  Our larger window is also known as the `Scutum Supercloud', so named for the anomalously high areal density of stars there. 
Both of these windows are used to infer properties of the stellar populations of the Galactic bar.  Constraining the attenuation as a function of distance, as we have done, will no doubt be important part of those studies. \\

\noindent
{\it M. Castellanos:} What about new constraints on the abundance of N? An increase in N/H by a factor of 2 relative to solar would explain the higher [\nii]/\ha\ values (1--1.5). Is this justified to assume that 'ad hoc'? \\

\noindent
{\it Madsen:}  Solar gas phase abundances are usually assumed in the analysis of [\nii]/\ha, as well as other line ratio analyses.  You are correct, that an accurate determination of abundances is needed to more accurately  constrain photoionization models.  However, estimates of abundances in the different environments we are probing here are difficult to obtain. Future work in this area will improve the accuracy of the line-ratio analysis. \\

\noindent
{\it A. Kutyrev:} How does the extinction at $l=25$\dg\ 'window' compare to lower latitude locations around $l=20$\dg? \\

\noindent
{\it Madsen:} The attenuation near $l=25$\dg\ appears to be isolated only to a 5\dg$\times$5\dg\ area, as shown in Figure 2. We do not see similar \ha\ emission components from the distant Scutum arm other than in this region.


\begin{chapthebibliography}{}

\bibitem{} Barnes, D.G. et al 2001, MNRAS, 322, 486
\bibitem{} Collins, J.A. \& Rand, R.J. 2001, ApJ, 551, 57
\bibitem{} Dennison, B., Simonetti, J.H. \& Topsana, G.A. 1998, PASA, 15, 147
\bibitem{} Hoyle, F, \& Ellis, G.R.A 1963, Austral. J. Phys., 16, 1
\bibitem{} Gaustad, J.E. et al 2001, PASP, 113, 1326
\bibitem{} Gu{\'e}lin, M. in IAU Symp. 60, Galactic Radio Astronomy, ed. F.J. Kerr \& S.C. Simonson (Boston: D. Reidel), 51
\bibitem{} Haffner, L.M., Reynolds, R.J., \& Tufte, S.L. 1999, ApJ, 523, 223
\bibitem{} Haffner, L.M. et al 2003, ApJS, in press
\bibitem{} Hartmann, D. \& Burton, W.B. 1997, Atlas of Galactic Neutral Hydrogen (New York: Cambridge Univ. Press)
\bibitem{} Heiles, C., Reach, W.T., \& Koo, B-C. 1996, ApJ, 466,191
\bibitem{} Madsen, G.J. et al 2001, ApJ, 560, L135
\bibitem{} McKee, C.F. \& Ostriker, J.P. 1977, ApJ, 218, 148
\bibitem{} Mille, W.W., \& Cox, D.P. 1993, ApJ, 417, 579
\bibitem{} Parker, Q.A. \& Phillipps, S. 1998, PASA, 15, 28
\bibitem{} Rand, R.J. 1997, ApJ, 474, 129
\bibitem{} Reynolds, R.J. 1989, ApJ, 345, 811
\bibitem{} Reynolds, R.J. 1990, ApJ, 349, L17
\bibitem{} Reynolds, R.J. 1991, ApJ, 372, L17
\bibitem{} Reynolds, R.J., Roesler, F.L., \& Scherb, F. 1974, ApJ, 192, L53
\bibitem{} Reynolds, R.J. \& Tufte, S.L. 1995, ApJ, 439, L17
\bibitem{} Reynolds, R.J. et al 1998, ApJ, 494, L99
\bibitem{} Reynolds, R.J., Haffner, L.M., \& Tufte, S.L. 1999, ApJ, 525, L21
\bibitem{} Reynolds, R.J. et al 2001, ApJ, 548, L221
\bibitem{} Spitzer, L.J. \& Fitzpatrick, E.L. 1993, ApJ, 409, 299
\bibitem{} Taylor, J.H. \& Cordes, J.M. 1993, ApJ, 411, 674

\end{chapthebibliography}
\end{document}